\documentclass[onepage,pra]{revtex4}

\usepackage{amsthm,amsmath}
\usepackage{graphicx}
\usepackage{amsfonts}
\usepackage{amssymb}
\usepackage{palatino}
\usepackage{hyperref}

\def\bra#1{\mathinner{\langle{#1}|}}
\def\ket#1{\mathinner{|{#1}\rangle}}

\let\protect\relax
{\catcode`\|=\active
  \xdef\Braket{\protect\expandafter\noexpand\csname Braket \endcsname}
  \expandafter\gdef\csname Braket \endcsname#1{\begingroup
     \ifx\SavedDoubleVert\relax
       \let\SavedDoubleVert\|\let\|\BraDoubleVert
     \fi
     \mathcode`\|32768\let|\BraVert
     \left\langle{#1}\right\rangle\endgroup}
}
\def\BraVert{\@ifnextchar|{\|\@gobble}
     {\egroup\,\mid@vertical\,\bgroup}}
\def\BraDoubleVert{\egroup\,\mid@dblvertical\,\bgroup}
\let\SavedDoubleVert\relax

{\catcode`\|=\active
  \xdef\set{\protect\expandafter\noexpand\csname set \endcsname}
  \expandafter\gdef\csname set \endcsname#1{\mathinner
        {\lbrace\,{\mathcode`\|32768\let|\midvert #1}\,\rbrace}}
  \xdef\Set{\protect\expandafter\noexpand\csname Set \endcsname}
  \expandafter\gdef\csname Set \endcsname#1{\left\{%
     \ifx\SavedDoubleVert\relax \let\SavedDoubleVert\|\fi
     \:{\let\|\SetDoubleVert
     \mathcode`\|32768\let|\SetVert
     #1}\:\right\}}
}
\def\midvert{\egroup\mid\bgroup}
\def\SetVert{\@ifnextchar|{\|\@gobble}
    {\egroup\;\mid@vertical\;\bgroup}}
\def\SetDoubleVert{\egroup\;\mid@dblvertical\;\bgroup}

\begingroup
 \edef\@tempa{\meaning\middle}
 \edef\@tempb{\string\middle}
\expandafter \endgroup \ifx\@tempa\@tempb
 \def\mid@vertical{\middle|}
 \def\mid@dblvertical{\middle\SavedDoubleVert}
\else
 \def\mid@vertical{\mskip1mu\vrule\mskip1mu}
 \def\mid@dblvertical{\mskip1mu\vrule\mskip2.5mu\vrule\mskip1mu}
\fi

\newtheorem{lemma}{Lemma}
\newtheorem{Theorem}{Theorem}

\newcounter{lem}

\def\claim#1{\par\medskip\noindent\refstepcounter{lem}\hbox{\bf A\arabic{lem}. #1.}
\it\ 
}
\def\endclaim{
\par\medskip}

\begin{document}

\title{\sf \bfseries Universal freezing of quantum correlations within the geometric approach}

\author{Marco Cianciaruso$^{1,2}$, Thomas R. Bromley$^{1}$, Wojciech Roga$^{2}$, \\ Rosario Lo Franco$^{1,3,4}$, and Gerardo Adesso$^{1,*}$}
\affiliation{\quad \\ $^{1}$School of Mathematical Sciences, The University of Nottingham, University Park, Nottingham NG7 2RD, United Kingdom
\\$^{2}$
Dipartimento di Ingegneria Industriale, Universit\`a degli Studi di Salerno, Via Giovanni Paolo II 132, I-84084 Fisciano (SA), Italy
\\$^3$Dipartimento di Fisica e Chimica, Universit\`{a} di Palermo, via Archirafi 36, I-90123 Palermo, Italy
\\$^4$Instituto de F{\'{i}}sica de S{\~{a}}o Carlos, Universidade de S{\~{a}}o Paulo, Caixa Postal 369, 13560-970 S{\~{a}}o Carlos, S{\~{a}}o Paulo, Brazil \\
$^*$Corresponding author: \texttt{gerardo.adesso@nottingham.ac.uk}}

\begin{abstract}
{\bf Quantum correlations in a composite system can be measured by resorting to a geometric approach, according to which the distance from the state of the system to a suitable set of classically correlated states is considered. Here we show that all distance functions, which respect natural assumptions of invariance under transposition, convexity, and contractivity under quantum channels, give rise to geometric quantifiers of quantum correlations which exhibit the peculiar {\em freezing} phenomenon, i.e., remain constant during the evolution of a paradigmatic class of states of two qubits each independently interacting with a non-dissipative decohering environment. Our results demonstrate from first principles that freezing of geometric quantum correlations is independent of the adopted distance and therefore universal. This finding paves the way to a deeper physical interpretation and future practical exploitation of the phenomenon for noisy quantum technologies.}
\end{abstract}

\maketitle

\title[Universality of the freezing of geometric quantum correlations]{}

\section*{\sf \bfseries INTRODUCTION}

\noindent {\Huge{I}}n quantum mechanics, the mathematical description of a composite quantum system is based on both the superposition principle and the tensorial structure of the Hilbert space associated with it. The coexistence of these two principles makes the properties of generic states of a composite quantum system particularly weird, in the sense that they cannot be reproduced by any state of a classical system. Some of the most striking non-classical properties exhibited by quantum states can be collected under the name of {\it quantum correlations} \cite{Horodecki2009,Modi2012}. Nowadays, there is universal consensus on the fact that the quantum correlations shared by two subsystems in a global pure state are entirely captured by entanglement and can be quantified by any valid entanglement measure \cite{Horodecki2009,Vidal2000}. On the other hand, it is also now quite clear that there exist non-entangled mixed states still manifesting some non-classical features, such as an unavoidable disturbance due to local measurements, which embodies the concept of quantum discord \cite{Henderson2001,Ollivier2001}. Therefore, in realistic open quantum systems, entanglement may represent only a portion, sometimes negligible, of the quantumness of correlations, while more general figures of merit to quantify quantum correlations are provided by suitable measures of discord-type correlations \cite{Modi2012,Aaronson2013a}. Despite an intense recent activity in investigating interpretation, quantification, and applications of discord and related quantifiers of quantum correlations \cite{Modi2012}, these quantities still remain less understood than entanglement.

In order to unveil the most profound signatures of quantumness in composite systems, it is essential to identify mathematically rigorous and physically meaningful properties that differentiate the notion of discord-type quantum correlations from that of entanglement (and of classical correlations), and are manifested by any valid measure thereof. One such property common to generic discord-type correlation measures is, for instance, the absence of monogamy \cite{arequantum}. Besides the fundamental implications, this area of investigation has a technological motivation \cite{merali,golden}, since quantum correlations beyond and even without entanglement have been shown to play a resource role for certain schemes of quantum computation \cite{datta,dqc1expwhite,dqc1explaflamme}, communication \cite{madhok:pr2011a,cavalcanti:pr2011a,np1,np2}, and metrology \cite{modix,lqu,interpower}. Finding valuable and general traits of these quantum correlation resources, in particular for what concerns their dynamical preservation during the unavoidable interaction of a principal quantum systems with the surrounding environments, constitutes an important aim with a clear potential to lead to useful recipes for their practical exploitation.

Numerous works have in fact investigated the dynamics of general quantum correlations in open quantum systems undergoing various types of Markovian or non-Markovian evolution, as reviewed e.g.~in \cite{Modi2012,revbraz,lofrancoreview}. Although different measures of quantum correlations can exhibit distinct features and impose inequivalent orderings on the set of quantum states, it has emerged as a general trait that  discord-type quantum correlations are more robust than entanglement against noise \cite{Streltsov2011,Ciccarello2012,maziero,acinferraro,lofrancosavasta} (see also \cite{stevecampbell,franchino} for a critical assessment) and cannot generally vanish at a finite evolution time (due to the fact that zero-discord states are of null measure \cite{acinferraro}), while entanglement can suffer so-called sudden death \cite{yu2009Science, almeida2007}. However, a fascinating and nontrivial phenomenon of {\it extreme} robustness to noise exhibited by general quantum correlations deserves special attention, and is the subject of our investigation.

Namely, under local non-dissipative decoherence evolutions, it has been observed that a number of known discord-type measures all remain constant (`frozen') for a finite time interval in Markovian conditions \cite{Aaronson2013a,Mazzola,necsuff}, and for multiple intervals \cite{Mazzola2011,LoFrancoAndersson2012,LoFranco2013} or forever \cite{Haikka2013} in non-Markovian conditions, when considering two non-interacting qubits initially in a specific class of Bell-diagonal states. This freezing phenomenon, not  exhibited by any measure of entanglement, is quite appealing since it implies that every protocol relying on discord-type quantum correlations as a resource will run with a performance unaffected by noise in the specific dynamical conditions.
Currently, the occurrence of freezing has been investigated by explicitly considering the evaluation of specific discord-type measures on a case by case basis \cite{Mazzola,Mazzola2011,Aaronson2013a}. However, it is natural to ask whether this phenomenon is a mere mathematical accident due to the particular choices of quantum correlations quantifiers, or whether it must manifest independently of the adopted measure, thus having a universal character and promising to bear a deep physical meaning.
This work addresses such an issue.

We prove that freezing occurs for any geometric measure of quantum correlations, whenever the distance defining the measure respects a minimal set of physical assumptions, namely dynamical contractivity under quantum channels, invariance under transposition, and convexity. The freezing phenomenon is therefore revealed as {\it universal} within the geometric approach to quantum correlations.
Notice that our work differs from other complementary investigations of the freezing phenomenon \cite{necsuff,HRI}. In particular, in a recent work \cite{HRI}, the authors provide necessary and sufficient conditions for a general state to exhibit freezing under non-dissipative decoherence, according to some specific measure of discord. Here, instead, we focus on a specific class of initial states, and we identify the minimal set of conditions that {\it any} general distance-based measure of discord needs to satisfy in order to freeze. On some random family of initial states, it is certainly possible to see freezing according to  one discord-type measure but not  to another. What we prove here  is that, for the specific class of Bell-diagonal states identified in \cite{Mazzola,Aaronson2013a}, all {\it bona fide} geometric quantifiers of quantum correlations (respecting the three physical assumptions mentioned above) undergo the same dynamics, featuring the freezing phenomenon. In proving the main result, we also introduce and characterise a global quantum control channel which can completely invert decoherence on a subset of Bell-diagonal two-qubit states, and can be of independent interest.

The paper is organised as follows. We first summarise the properties any valid measure of quantum correlations is expected to hold, and provide general definitions for distance-based measures, distinguishing between those for entanglement and those for discord-type correlations. In the {\bf Results} section,  we illustrate the freezing phenomenon in geometric terms, by considering for convenience the specific case of the Bures distance-based measure of quantum correlations. We then present the main result, by proving that freezing must happen for any bona fide distance-based discord-type measure. Finally, we offer our conclusions in the {\bf Discussion} section. Some technical bits are deferred to  the {\bf Methods} section.

\medskip

{\bf Measures of quantum correlations}.
 Here we recall the requirements that a valid measure of quantum correlations is expected to have, later focusing specifically on geometric (distance-based) definitions. For more details on the quantification of quantum correlations, the reader is referred to recent review articles \cite{Horodecki2009,Modi2012}.

In this paper, we consider a state $\rho \equiv \rho^{AB}$ of a two-qubit system, with subsystems tagged  $A$ and $B$. As is well known \cite{Werner89,Horodecki2009}, entanglement quantifiers capture the degree of non-separability of the state $\rho$ of the global system; the corresponding distance-based measures of entanglement are calculated from the set $\mathcal{S}$ of separable states \cite{Vedral1997}, i.e., states $\sigma$ which can be written as convex combination of product states,
\begin{equation}
\label{rosecco}
\sigma = \sum_i p_i \varrho_i^A \otimes \varpi_i^B\,,
\end{equation}
where $p_i$ is a probability distribution, while $\{\varrho_i^A\}$ and $\{\varpi_i^B\}$ are arbitrary ensembles of states for subsystem $A$ and $B$, respectively.

 Quantifiers of quantum correlations other than entanglement, the so-called discord-type measures, capture instead the minimal degree of disturbance on the state $\rho$ after local projective measurements on the system \cite{Ollivier2001,Henderson2001,Modi2012}. The projective measurements can be performed either on a subsystem only (which gives rise to one-way, asymmetric discord-type measures) or on both subsystems (which gives rise to two-way, symmetric discord-type measures). These two versions of discord-type measures have valuable operational meanings in different contexts \cite{Modi2012} and the corresponding distance-based measures are calculated, respectively, from the set of classical-quantum (CQ) and classical-classical (CC) states \cite{PianiBroad}. Explicitly, a CQ state (with respect to measurements on subsystem $A$) is a particular type of separable state, which can be written in the form
\begin{equation}
\label{romedio}
\chi = \sum_i p_i \ket{i}\bra{i}^A \otimes \varpi_i^B\,,
\end{equation}
 where $p_i$ is a probability distribution,  $\{\ket{i}^A\}$ denotes an orthonormal basis for subsystem $A$, and $\{\varpi_i^B\}$ is an arbitrary ensemble of states for subsystem $B$.  Similarly, a CC state can be written in the form
\begin{equation}
\label{rochiatto}
\chi = \sum_{i,j} p_{i,j} \ket{i}\bra{i}^A \otimes \ket{j}\bra{j}^B\,,
\end{equation}
where $p_{ij}$ is a joint probability distribution, while  $\{\ket{i}^A\}$ and $\{\ket{j}^B\}$ denote orthonormal bases for subsystem $A$ and $B$, respectively. Clearly, the set $\mathcal{C}$ of CC states is contained in the set of CQ states, which is a subset of the set $\mathcal{S}$ of separable states.

In general, if not explicitly written, by ``classical'' states we hereafter mean CC states, i.e., states $\chi$ which are diagonal in a product basis, as defined by Eq.~(\ref{rochiatto}); these states correspond merely to the embedding of a bipartite probability distribution $\{p_{ij}\}$ into the quantum formalism.   We further specify that for Bell-diagonal states \cite{HoroBell}, namely the specific class of two-qubit states considered in this work, the two notions of discord are completely equivalent, therefore our conclusions about the universality of the freezing will apply indifferently to both one-way and two-way geometric measures of discord-type correlations.

From a quantitative point of view, a valid entanglement measure, also known as an entanglement monotone \cite{Vidal2000}, is any real and nonnegative function $E$ on the set of states $\rho$ satisfying the following basic axioms \cite{Horodecki2009, PlenioVirmani}:
\begin{enumerate}
\item[(E.i)] $E(\rho)=0$ if $\rho$ is a separable state as defined in Eq.~(\ref{rosecco});
\item[(E.ii)] $E$ is invariant under local unitaries, i.e. $E \left((U_A\otimes U_B) \rho (U_A^\dagger \otimes U_B^\dagger) \right)=E(\rho)$ for any state $\rho$ and any local unitary operation $U_A$
$(U_B)$ acting on subsystem $A$ ($B$);
\item[(E.iii)] $E$ is monotonically nonincreasing under local operations and classical communication (LOCC), i.e. $E\big(\Lambda_{\rm LOCC}(\rho)\big)\leq E(\rho)$ for any state $\rho$ and any LOCC channel
$\Lambda_{\rm LOCC}$\,.
\end{enumerate}
Furthermore, additional  properties for an entanglement measure can include convexity,
\begin{enumerate}
\item[(E.iv)] $E$ is convex, i.e. $E(q\rho + (1-q)\sigma)\leq q E(\rho) + (1-q)E(\sigma)$, for any pair of states $\rho,\sigma$ and any $q\in[0,1]$.
\end{enumerate}
Notice, however, that while convexity is physically desirable (as it would mean that entanglement cannnot increase by mixing states), it is not an essential property, since there are valid entanglement monotones which are not convex \cite{PlenioLogNeg}.

The theory of quantum correlations other than entanglement is not completely developed yet \cite{Modi2012,ABC}, but we can nonetheless identify some {\it desiderata} for any quantifier thereof. A (two-way) discord-type measure is any real and nonnegative function  $Q$ on the set of states $\rho$ satisfying the following requirements:
\begin{enumerate}
\item[(Q.i)] $Q(\rho)=0$ if $\rho$ is a classical state as defined in Eq.~(\ref{rochiatto});
\item[(Q.ii)] $Q$ is invariant under local unitaries, i.e. $Q \left((U_A\otimes U_B) \rho (U_A^\dagger \otimes U_B^\dagger) \right)=Q(\rho)$ for any state $\rho$ and any local unitary operation $U_A$ $(U_B)$ acting on subsystem $A$ ($B$);
\item[(Q.iii)] $Q$ is monotonically nonincreasing under local commutativity preserving quantum channels $\Lambda_A\otimes \Lambda_B$, i.e. $Q \big((\Lambda_A\otimes \Lambda_B)(\rho)\big) \leq Q(\rho)$ for any state $\rho$ and any commutativity preserving map $\Lambda_{A(B)}$ on subsystem $A(B)$, that is, $[\Lambda_{A(B)}(\rho_{A(B)}),\Lambda_{A(B)}(\sigma_{A(B)})]=0$ when $[\rho_{A(B)},\sigma_{A(B)}]=0$ for arbitrary marginal states $\rho_{A(B)}$ and $\sigma_{A(B)}$;
\item[(Q.iv)] $Q$ reduces to an entanglement measure for pure states, i.e. $Q(|\psi^{AB}\rangle)=E(|\psi^{AB}\rangle)$ for any pure state $|\psi^{AB}\rangle$.
\end{enumerate}
We introduce property (Q.iii) in analogy with property (E.iii) for entanglement. Namely,  it is known that local commutativity preserving channels cannot create discord-type correlations, as they leave the set of classical states invariant \cite{Xueyuan2012}. We thus require that any valid measure of discord-type correlations should be monotonically nonincreasing under such channels. Notice, in particular, that for two qubits these channels include local unital channels \cite{Streltsov2011}.

The above requirements need to be slightly modified if a one-way discord-type measure $Q^\rightarrow$, say with measurements on $A$, is considered. Specifically, property (Q.i) becomes: $Q^{\rightarrow}(\rho)=0$ if $\rho$ is a CQ state as defined in Eq.~(\ref{romedio}). Furthermore, a stricter monotonicity requirement supplements (Q.iii) for all valid one-way discord-type measures \cite{Streltsov,PianiAdesso,Aaronson2013a}, namely
\begin{enumerate}
\item[(Q.iii.bis)]  $Q^\rightarrow$ is monotonically nonincreasing under arbitrary local quantum channels on the unmeasured subsystem $B$, that is, $Q^\rightarrow \big((\mathbb{I}_A\otimes \Lambda_B)(\rho)\big) \leq Q^\rightarrow(\rho)$ for any state $\rho$ and any completely positive trace-preserving (CPTP) map $\Lambda_{B}$ on subsystem $B$.
\end{enumerate}
Properties (Q.ii) and (Q.iv) apply equally to two-way and one-way discord-type measures. The latter property just signifies that, in pure bipartite states, there is a unique kind of quantum correlations, arising in all but tensor product states. Even correlations  stronger than entanglement, such as steering and nonlocality, just collapse back to non-separability in the case of pure states.

In order to investigate the freezing phenomenon, in this paper we resort to a {\it geometric} approach to define a very general class of valid measures of quantum correlations. According to such an approach, the entanglement $E$ and the discord-type correlations $Q$ of a state $\rho$ can be quantified as the minimal distance from $\rho$ to the sets $\mathcal{S}$ and $\mathcal{C}$ of separable and classical states, respectively \cite{Vedral1997,Dakic2010}. In formulae,
\begin{eqnarray} \label{eq:geometricentanglement}
E_D(\rho)\equiv\inf_{\sigma\in \mathcal{S}}D(\rho,\sigma),\\
Q_D(\rho)\equiv\inf_{\chi\in \mathcal{C}}D(\rho,\chi), \label{eq:geometricquantumcorrelations}
\end{eqnarray}
where separable states $\sigma$ are defined by Eq.~(\ref{rosecco}), while classical states $\chi$  are defined respectively by Eqs.~(\ref{rochiatto}) and (\ref{romedio}) depending on whether a two-way or one-way discord-type measure is considered. In these definitions,  $D$ can denote in principle any suitable distance on the set of quantum states.

In order for the geometric measures $E_D$ and $Q_D$ to respect the essential properties listed above, the distance $D$ needs to satisfy certain mathematical requirements \cite{nielsenchuang}.
Here, we identify a minimal set of three such requirements, that will be said to characterise $D$ as a bona fide distance. Given any  states $\rho$,  $\sigma$, $\tau$, and $\varsigma$, these are:
\begin{enumerate}
\item[(D.i)] Contractivity under CPTP maps, i.e.
\begin{equation}\label{eq:contractivity}
D(\Lambda(\rho),\Lambda(\sigma))\leq D(\rho,\sigma),\\
\end{equation}
for any CPTP map $\Lambda$;
\item[(D.ii)] Invariance under transposition
\begin{equation}\label{eq:antiunitarity}
D(\rho^T, \sigma^T) = D(\rho, \sigma)\,;
\end{equation}
\item[(D.iii)] Joint convexity, i.e.
\begin{equation}\label{eq:convexity}
D\big(q \rho+(1-q)\sigma, q \tau + (1-q)\varsigma\big)\leq q D(\rho,\tau) + (1-q)D(\sigma,\varsigma)
\end{equation}
for any  $q\in[0,1]$.
\end{enumerate}
Let us comment on the physical significance of these requirements.

On the one hand, the contractivity property (D.i) of $D$, Eq.~(\ref{eq:contractivity}) is a fundamental requirement for a distance in quantum information theory \cite{nielsenchuang}, and has two purposes. First, it makes $D$ a statistically relevant distance, due to the fact that non-invertible CPTP maps are the mathematical counterparts of noise and the latter cannot lead to any increase in the information related to the distinguishability of quantum states  \cite{Bengtsson2006,Rivas2014}. Second, (D.i)  makes the distance-based measures of entanglement $E_D$ and quantum correlations $Q_D$ [Eqs.~(\ref{eq:geometricentanglement}) and (\ref{eq:geometricquantumcorrelations})] physically meaningful, by implying the essential properties (E.ii), (E.iii), (Q.ii), (Q.iii), and also (Q.iii.bis), where the latter applies to the corresponding distance-based one-way discord-type measure $Q^{\rightarrow}_D$ defined by choosing the set of CQ (rather than CC) states in Eq.~(\ref{eq:geometricquantumcorrelations}). Notice that (D.i) implies in particular the standard property of invariance of the distance $D$ under unitary operations, since they can be seen as reversible CPTP maps; namely, $D(U \rho U^{\dagger}, U \sigma U^{\dagger})=D(\rho,\sigma)$, for any pair of states $\rho$, $\sigma$, and any unitary $U$.

On the other hand, the invariance of a distance $D$ under transposition, Eq.~(\ref{eq:antiunitarity}), is not typically discussed in the literature. However,
transposition of an $N\times N$ hermitian matrix, which amounts to complex conjugation in the computational basis, corresponds to a reflection in a $[N(N+1)/2-1]$-dimensional hyperplane. Property (D.ii) thus means that a distance $D$ on the set of quantum states is assumed to be invariant under reflections, which appears as a fairly natural requirement \cite{Bengtsson2006}. Notice that, together with (D.i), this property implies invariance of the distance $D$ under antiunitary operations. Any antiunitary matrix $O$ can be expressed as $O=UC$, where $U$ is a unitary matrix and $C$ denotes complex conjugation in the computational basis. We have then  $D(O\rho O^{-1},O\sigma O^{-1}) = D\big(UC(\rho)U^\dagger,UC(\sigma)U^\dagger\big) = D\big(C(\rho),C(\sigma)\big) = D(\rho^T,\sigma^T)=D(\rho,\sigma)$, for any pair of states $\rho$, $\sigma$.

Finally, the joint convexity property (D.iii) of $D$, Eq.~(\ref{eq:convexity}), is also quite intuitive and it makes the corresponding distance-based entanglement measure $E_D$ convex, implying the desirable property (E.iv). Notice however that discord-type measures are, by contrast, neither convex nor concave \cite{LangCaves}, as the set $\mathcal{C}$ of classical states is not a convex set \cite{acinferraro}.

There are a number of known distances which satisfy the three physical assumptions listed above \cite{Bengtsson2006}, and have been employed to define valid geometric measures of quantum correlations.  Suitable examples for $D$ include in particular the the relative entropy \cite{Vedral1997,Modi2010}, the squared Bures distance \cite{Vedral1998,Spehner2013,Aaronson2013a}, the squared Hellinger distance \cite{LuoHell,HanggiHell,lqu}, and the trace (or Schatten one-norm) distance \cite{Eisert2003,Nakano2013,Sarandy2013}. Contrarily, the Hilbert-Schmidt distance does not respect the contractivity property (D.i), and earlier attempts to adopt it to define geometric measures of quantum correlations \cite{Witte1999,Dakic2010}  have led to inconsistencies \cite{Ozawa2000,PianiComment}.

In this paper, we label a generic distance $D$ obeying properties (D.i), (D.ii), and (D.iii) as a bona fide one, and the associated distance-based quantities $E_D$ and $Q_D$ as bona fide measures of entanglement and discord-type correlations respectively. Therefore, the main result of this paper will be a proof of the universality of the freezing for all geometric measures of quantum correlations constructed via bona fide distances as formalised in this Section.

\section*{\sf \bfseries RESULTS}

{\bf Freezing of quantum correlations measured by Bures distance}.
We now present the freezing phenomenon from a geometric perspective, by employing a particular bona fide measure of quantum correlations, that is the Bures distance-based measure \cite{Spehner2013,Spehner2014,Aaronson2013a,Bromley2014}. We first recall all the basic ingredients for the complete description of the phenomenon.

The Bures distance $D_{\rm Bu}$ between two states $\rho$ and $\sigma$ is defined as
\begin{equation}\label{eq:Buresdistance}
D_{\rm Bu}(\rho, \sigma) = \sqrt{2\left(1 - \sqrt{F(\rho,\sigma)}\right)},
\end{equation}
where
\begin{equation}\label{Eq:Fidelity}
F(\rho, \sigma) = \left(\mbox{Tr}\left[ \sqrt{\sqrt{\rho} \sigma \sqrt{\rho}}\right]\right)^2,
\end{equation}
is the Uhlmann fidelity \cite{uhlmann}.

The Bures distance arises from a specific case of a general family of Riemannian contractive metrics on the set of density matrices, characterised by Petz \cite{Petz1996} following the work by Morozova and \u{C}encov \cite{Morozova1991}. It can be connected operationally to the success probability in ambiguous quantum state discrimination, and it has been successfully employed to define geometric measures of entanglement, quantum, classical, and total correlations \cite{Streltsov2010,Streltsov2011,Aaronson2013a,Spehner2013,Spehner2014,Roga2014,Bromley2014}.
Specifically, the Bures entanglement measure $E_{\rm Bu}$ and the Bures discord-type measure $Q_{\rm Bu}$ are defined by choosing in Eqs.~(\ref{eq:geometricentanglement}) and (\ref{eq:geometricquantumcorrelations}) the squared Bures distance $D \equiv D^{2}_{\rm Bu}$, which is a bona fide one obeying properties (D.i), (D.ii), and (D.iii).

Bell-diagonal (BD) states, also referred to as T-states or two-qubit states with maximally mixed marginals \cite{HoroBell},  are structurally simple states which nonetheless remain of high relevance to theoretical and experimental research in quantum information, as they include the well-known Bell and Werner states \cite{Werner89} and can be employed as resources for operational tasks such as entanglement activation and distribution via discord-type correlations \cite{PianiAdesso,Fedrizzi2013,Sciarrino,kay2012}. BD states are by definition diagonal in the basis of the four maximally entangled Bell states, and their Bloch representation in the computational basis is
\begin{equation}\label{BDstate}
\rho = \frac{1}{4} \left(\mathbb{I}^{A} \otimes \mathbb{I}^{B} + \sum_{i=1}^{3} c_i \sigma_{i}^{A} \otimes \sigma_{i}^{B} \right),
\end{equation}
where $c_i = \mbox{Tr} \left[ \rho \left( \sigma_{i}^{A} \otimes \sigma_{i}^{B} \right)\right]$, $\mathbb{I}$ is the identity matrix and $\{\sigma_{i}\}$ are the Pauli matrices. Because $\rho$ is positive semidefinite, the vector $\vec{c} \equiv \{c_1,c_2,c_3\}$ characterising any BD state $\rho$ is constrained to lie within the tetrahedron with vertices $\{1,-1,1\}$, $\{-1,1,1\}$,
$\{1,1,-1\}$ and $\{-1,-1,-1\}$ \cite{HoroBell,Dakic2010,LangCaves}; the vertices represent respectively the four Bell states
\begin{eqnarray}\label{eq:BellStates}
|\Phi^{+}\rangle &=& \frac{1}{\sqrt2} (\ket{00} + \ket{11})\,,\nonumber \\
|\Phi^{-}\rangle &=& \frac{1}{\sqrt2} (\ket{00} - \ket{11})\,, \nonumber \\
& & \\
|\Psi^{+}\rangle &=& \frac{1}{\sqrt2} (\ket{01} + \ket{10})\,, \nonumber \\
|\Psi^{-}\rangle &=& \frac{1}{\sqrt2} (\ket{01} - \ket{10})\,. \nonumber
\end{eqnarray}
For convenience, in the following, we shall specify an arbitrary BD state $\rho$ by referring equivalently to its defining triple
$\{c_1,c_2,c_3\}$.

A non-dissipative quantum channel acting on a qubit induces decoherence with no excitation exchange between the qubit and its environment. We consider the evolution of two non-interacting qubits undergoing local identical non-dissipative decoherence channels. The action of any such channel on each single qubit is characterised by the following Lindblad operator \cite{Mazzola,Aaronson2013a},
\begin{equation}\label{evolvc}
\mathcal{L}_{k} [\rho_{m}] = \frac{\gamma}{2} \left( \sigma_{k}^{m} \rho_{m}\sigma_{k}^{m} - \rho_{m} \right),
\end{equation}
where $\gamma \geq 0$ is the decoherence rate, $\rho_m$ is the reduced state of subsystem $m$ ($m=A,B$) and $k \in \{1,2,3\}$ represents the direction of the noise. Namely, the choice of $k =1,2,3$ respectively identifies decoherence in the Pauli $x, y, z$ basis for each qubit, and the corresponding channels are known in the quantum computing language as bit flip ($k=1$), bit-phase flip ($k=2$), and phase flip ($k=3$, also known as pure dephasing) channels \cite{nielsenchuang,Mazzola}. It is worth noting that one can easily derive the dynamics of the composite two-qubit system from the dynamics of the single qubits, since each of the two qubits is locally interacting only with its own environments (and not with the other qubit), so that they have independent dynamical evolutions \cite{lofrancoreview}.

Equivalently, the evolution of a two-qubit state $\rho$ under local non-dissipative decoherence channels can be obtained in the operator-sum representation by the map
\begin{equation}\label{eq:dynamicalevolutionunderlocalidenticalindependentpuredephasing} 
\Lambda^t_{{\rm LD}_k}(\rho)=\sum_{i,j=1}^4 K_i^A\otimes K_j^B \rho K_i^{A \dagger}\otimes K_j^{B \dagger},
\end{equation}
where the single-qubit Kraus operators are
\begin{equation}
K^m_k = \sqrt{\frac{1 - e^{- \gamma t}}{2}} \sigma_{k},\
K^m_{i,j\neq k}=0, \
K^m_4 = \sqrt{\frac{1 + e^{- \gamma t}}{2}} \mathbb{I},
\end{equation}
with $m=A,B$ and $k$ being the direction of the noise as in Eq.~(\ref{evolvc}).
Interestingly, from Eq.~(\ref{eq:dynamicalevolutionunderlocalidenticalindependentpuredephasing}) one can easily see that two non-interacting qubits initially in a BD state, undergoing local identical non-dissipative channels, maintain the BD structure for all time. More specifically, the triple $\vec{c}(t)$ characterizing the BD evolved state $\rho(t)$ can be written as follows
\begin{equation}\label{eq:timeevolvedBDstateunderlocalpuredephasing}
c_{i,j\neq k}(t)=c_{i,j}(0)e^{-2\gamma t},\ \ \ \ c_k(t)=c_k(0),
\end{equation}
where  $\vec{c}(0)$ is the triple characterising the initial BD state $\rho(0)$.

\begin{figure}[t]
    \centering
    \includegraphics[width=0.7\textwidth]{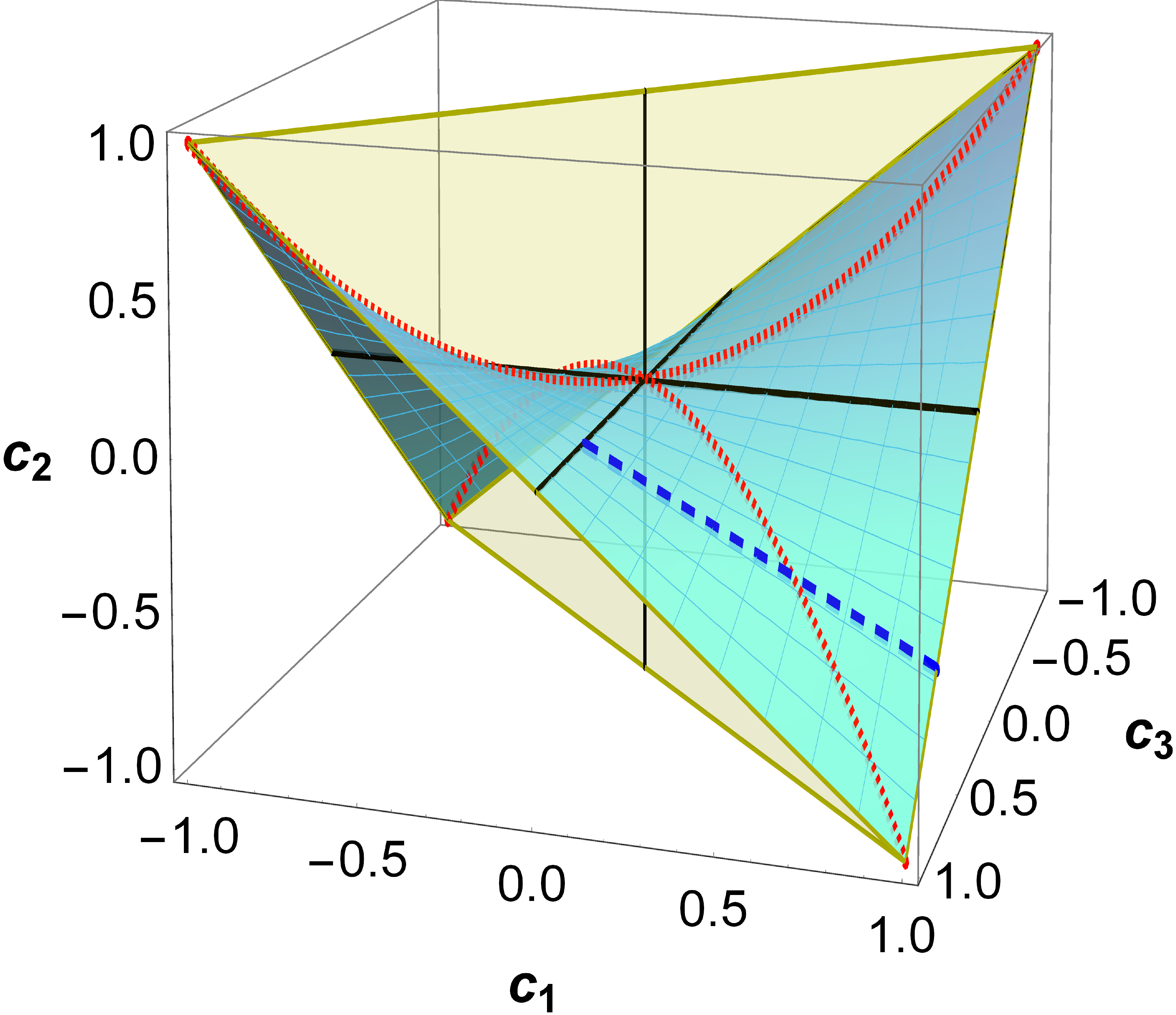}
    \caption{The phase flip freezing surface (meshed cyan) within the tetrahedron of all BD states (light yellow) represented in the  $\{c_{1},c_{2},c_{3}\}$ space. The surface contains all and only the BD states with triple $\{c_1,-c_1c_3,c_3\}$, and thus accommodates all the BD states respecting Eq.~(\ref{eq:constraintdefiningthefreezingsurface}). Solid black lines represent the classical BD states, which lie on the axes. The dotted red lines represent the threshold points on the surface when $|c_{1}| = |c_{3}|$, which occurs at the time $t=t^*$ defined in Eq.~(\ref{eq:thresholdtime}). For any state obeying the initial conditions of Eq.~(\ref{eq:constraintdefiningthefreezingsurface}), we show that  bona fide discord-type quantum correlations are frozen under local phase flip channels up to the time $t^*$. As an example, the dashed blue line represents the dynamical trajectory of the initial BD state $\{1,-0.6,0.6\}$, which evolves  under local phase flip channels moving towards the $c_{3}$-axis with increasing time; the discord-type correlations are frozen in the initial segment of the trajectory up to the intersection with the red dotted line, and decay exponentially afterwards, as plotted in Fig.~\ref{Fig:SpecificFreezing}.}
    \label{fig:thephaseflipfreezingsurface}
\end{figure}

For non-interacting qubits initially in a BD state $\rho(0)$, undergoing identical local non-dissipative channels, the freezing phenomenon for discord-type quantum correlations occurs if specific initial state conditions are satisfied.
For convenience and without loss of generality, from now on we focus our analysis on the phase flip (pure dephasing) channel ($k=3$), for which these initial conditions consist of the triples
$\vec{c}(0)$ such that
\begin{equation}\label{eq:constraintdefiningthefreezingsurface}
c_2(0)=-c_1(0)c_3(0),\quad
|c_1(0)|>|c_3(0)|.
\end{equation}
The suitable initial conditions for bit flip ($k=1$) or bit-phase flip ($k=2$) channels can be obtained by setting analogous relations among the coefficients $c_{1},c_{2},c_{3}$ \cite{Aaronson2013a}.
The BD states satisfying the constraint of Eq.~(\ref{eq:constraintdefiningthefreezingsurface}) distribute within a two-dimensional surface inside the tetrahedron of all BD states, which is shown in Fig.~\ref{fig:thephaseflipfreezingsurface} and will be referred to herein as (phase flip) freezing surface.

From Eq.~(\ref{eq:timeevolvedBDstateunderlocalpuredephasing}) one can see that the time evolved state $\rho(t)$ is a BD state characterised by the triple
$\{c_1(0)e^{-2\gamma t},-c_1(0)c_3(0)e^{-2\gamma t},c_3(0) \}$, which means that it remains confined within the freezing surface at any time. An example of this dynamical trajectory is represented in Fig.~\ref{fig:thephaseflipfreezingsurface} by the dashed blue line.

\begin{figure}[t]
    \centering
    \includegraphics[width=0.75\textwidth]{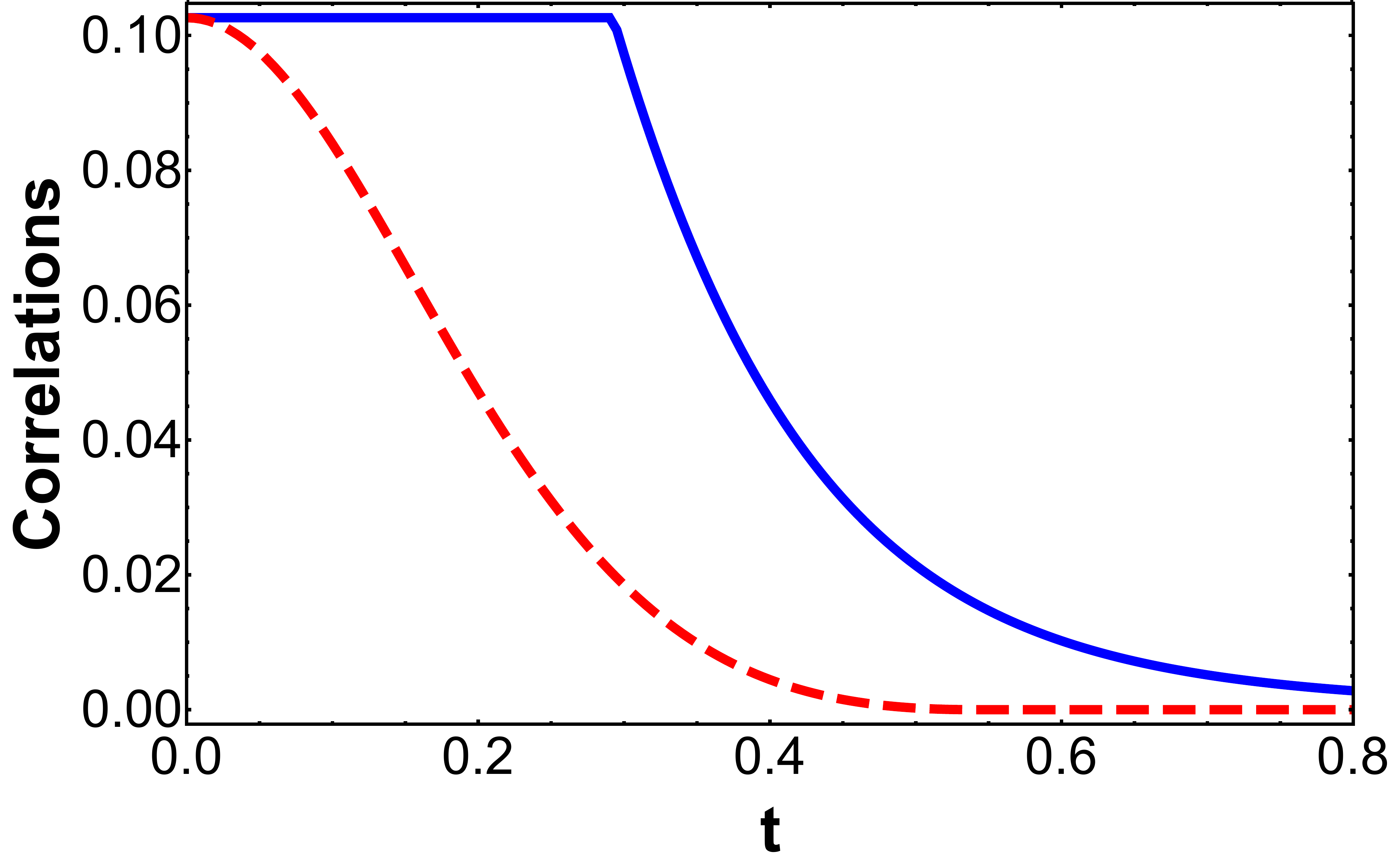}
    \caption{The paradigmatic freezing of Bures distance-based quantum correlations for an initial BD state of the form $\vec{c}(0) = \{1,-0.6,0.6\}$. The solid blue line represents the time evolution of discord-type correlations and the dashed red line represents the time evolution of Bures distance-based entanglement.}
    \label{Fig:SpecificFreezing}
\end{figure}

The above dynamical conditions give rise to a very peculiar evolution of the quantum correlations present in the time evolved state. Namely, defining a threshold time $t^*$ by
\begin{equation}\label{eq:thresholdtime}
 t^*\equiv -\frac{1}{2\gamma}\ln\frac{|c_3(0)|}{|c_1(0)|},
\end{equation}
we find that the Bures distance-based measure of discord $Q_{\rm Bu}$ stays constant (freezes) for $0 \leq t < t^*$ and then decays exponentially from $t>t^*$ onwards, as is shown in Fig.~\ref{Fig:SpecificFreezing}. This can be straightforwardly shown by exploiting the available closed formula for Bures discord-type correlations of BD states \cite{Aaronson2013a,Spehner2013}. On the contrary, entanglement measured e.g.~by $E_{\rm Bu}$ undergoes a typical sudden death at a finite time \cite{yu2009Science}.
We stress again that this behaviour of quantum correlations, here illustrated for $Q_{\rm Bu}$, has been independently observed (on a case by case basis) for several valid discord-type measures in the aforementioned dynamical conditions \cite{Aaronson2013a}: this paper will provide a rigorous basis to establish its universality within the bona fide geometric approach.

The freezing phenomenon can be understood in geometric terms by looking at Fig.~\ref{fig:thephaseflipfreezingsurface}, which represents the phase flip freezing surface containing BD states of the form $\{c_1,-c_1 c_3,c_3\}$, thus containing all the states with initial conditions identified in Eq.~(\ref{eq:constraintdefiningthefreezingsurface}). The solid black lines represent the axes in the $\{c_{1},c_{2},c_{3}\}$ space, which correspond to all the classical BD states. The dashed blue line represents a particular state evolution under local phase flip channels, leading to the freezing phenomenon. The intersection between the dotted red line and the dashed blue line represents the time $t^*$ such that $|c_1(t^*)|=|c_3(t^*)|$, i.e., the threshold time (\ref{eq:thresholdtime}) corresponding to the sudden change from freezing to decaying of quantum correlations.

For $t<t^*$, i.e.~when $|c_1(t)|>|c_3(t)|$, one of the closest classical states to the evolved BD state $\rho(t)=\{c_1(0)e^{-2\gamma t},-c_1(0)c_3(0)e^{-2\gamma t},c_3(0) \}$ is its Euclidean orthogonal projection onto the $c_1$-axis, i.e.~the BD classical state with triple $\chi_{\rho(t<t^*)}=\{c_1(0)e^{-2\gamma t},0,0\}$. From Eq.~(\ref{eq:geometricquantumcorrelations}) we know that the discord-type quantum correlations of $\rho(t)$ for $t<t^*$ are given exactly by the (squared) Bures distance between the evolving state $\rho(t<t^*)$ and this closest classical state $\chi_{\rho(t<t^*)}$. Interestingly, one can observe that this distance is constant for any $t<t^*$, which indeed implies that the quantum correlations of $\rho(t)$ are frozen for any $t<t^*$, given arbitrary initial conditions on the freezing surface defined by (\ref{eq:constraintdefiningthefreezingsurface}).

For $t>t^*$, i.e.~when $|c_3(t)|>|c_1(t)|$, one of the closest classical states to the evolved BD state $\rho(t)=\{c_1(0)e^{-2\gamma t},-c_1(0)c_3(0)e^{-2\gamma t},c_3(0) \}$ is instead its Euclidean orthogonal projection onto the $c_3$-axis, i.e.~the BD classical state with triple $\chi_{\rho(t>t^*)}=\{0,0,c_3(0)\}$, which is independent of time. Therefore the quantum correlations of the evolved state $\rho(t)$ decrease for any $t>t^*$, as the distance between the evolving state $\rho(t>t^*)$ and the steady closest classical state $\chi_{\rho(t>t^*)}$ decreases for any $t>t^*$.

For two-qubit BD states of the form (\ref{eq:constraintdefiningthefreezingsurface}) undergoing local pure dephasing evolutions, freezing (up to a time $t^*$) of geometric quantum correlations measured by the Bures distance thus relies on the following two properties:
\begin{enumerate}
\item[(F.i)] (according to the Bures distance)
one of the closest classical states to the evolved BD state $\rho(t)=\{c_1(t),-c_1(t)c_3(t),c_3(t)\}$ is the classical BD state $\{c_1(t),0,0\}$ when $|c_1(t)|>|c_3(t)|$;
\item[(F.ii)] the (Bures)
distance between the BD states with triples $\{c_1,-c_1c_2,c_3\}$ and $\{c_1,0,0\}$ does not depend on $c_1$, that is
\begin{equation}\label{eq:theweaktranslationalinvariance}
D(\{c_1,-c_1c_3,c_3\},\{c_1,0,0\})=D(\{0,0,c_3\},\{0,0,0\}),\  \forall c_1,c_3.
\end{equation}
\end{enumerate}

The main result of this paper will be to show that these two properties are satisfied by {\it any} contractive, transposition invariant, and convex distance, thus implying the freezing phenomenon for any bona fide distance-based measure of discord-type quantum correlations as defined above. Contrarily, we remark that the non-contractive Hilbert-Schmidt distance satisfies only the first property (F.i), whereas it does not manifest the kind of translational invariance expressed in Eq.~(\ref{eq:theweaktranslationalinvariance}), due to the fact that the trajectory of the evolved state is not parallel to the $c_1$-axis according to the Euclidean geometry, as is shown in Fig.~\ref{fig:thephaseflipfreezingsurface}. As a result, the Hilbert-Schmidt geometric discord \cite{Dakic2010}, which is not a bona fide measure \cite{PianiComment}, does not manifest freezing in the considered dynamical conditions, as previously observed \cite{Aaronson2013a}.

\medskip

{\bf Freezing of quantum correlations for all {\em bona fide} geometric measures}.
We now prove that for any distance $D$ respecting the bona fide requirements (D.i), (D.ii), and (D.iii), the aforementioned freezing properties  (F.i) and (F.ii) are satisfied.

Let us begin by proving that (D.i) $\Rightarrow$ (F.ii), i.e., that the very peculiar invariance on the phase flip freezing surface expressed in Eq.~(\ref{eq:theweaktranslationalinvariance}) follows from the contractivity of a distance $D$.
\begin{Theorem}\label{th:anycontractivedistanceismagictranslationalinvariant}
Any contractive distance $D$ satisfies the following translational invariance properties
\begin{equation}\label{eq:magictranslationalinvariance1}
D(\{c_1,-c_1c_3,c_3\},\{c_1,0,0\}) = D(\{0,0,c_3\},\{0,0,0\})\ \ \ \ \forall c_1,c_3
\end{equation}
and
\begin{equation}\label{eq:magictranslationalinvariance2}
D(\{c_1,-c_1c_3,c_3\},\{0,0,c_3\}) = D(\{c_1,0,0\},\{0,0,0\})\ \ \ \ \forall c_1,c_3
\end{equation}
where $\{c_1,-c_1c_3,c_3\}$ denotes any BD state on the phase flip freezing surface.
\end{Theorem}

\noindent \textit{Proof}. We first prove Eq.~(\ref{eq:magictranslationalinvariance1}). Considering an evolution due to  local complete pure dephasing channels acting on the two qubits, corresponding to the dynamical map $\Lambda^{\infty}_{{\rm LD}_3}$
as given in Eq.~(\ref{eq:dynamicalevolutionunderlocalidenticalindependentpuredephasing}) for $k=3$ and $t \rightarrow \infty$,
we get the inequality
\begin{eqnarray} \label{eq:inequalityduetodephasing1}
D(\{0,0,c_3\},\{0,0,0\})
&=& D(\Lambda^{\infty}_{{\rm LD}_3}\{c_1,-c_1 c_3,c_3\},\Lambda^{\infty}_{{\rm LD}_3}\{c_1,0,0\})  \nonumber \\
&\leq& D(\{c_1,-c_1 c_3,c_3\},\{c_1,0,0\}),
\end{eqnarray}
where the first equality is due to the fact that
\begin{equation}
\{0,0,c_3\} = \Lambda^{\infty}_{{\rm LD}_3} \{c_1,-c_1 c_3,c_3\},\quad
\{0,0,0\} = \Lambda^{\infty}_{{\rm LD}_3}\{c_1,0,0\},
\end{equation}
and the final inequality is due to contractivity of the distance $D$.

We now introduce a global two-qubit {\it rephasing} channel $\Lambda_{{\rm GR}_3}^{q}$ with operator-sum representation
\begin{equation}\label{Eq:KrausRepresentation}
\Lambda_{{\rm GR}_3}^{q}(\rho) = \sum_{i,\pm} K_{i,\pm}\rho K_{i,\pm}^{\dagger}
\end{equation}
where
\begin{eqnarray}
K_{1,\pm} = \sqrt{\frac{1\pm q}{2}} | \Phi^{\pm}\rangle \langle 00 |, \hspace{0.5 cm}& K_{2,\pm}=\sqrt{\frac{1\pm q}{2}} | \Psi^{\pm}\rangle \langle 01|, \nonumber \\
K_{3,\pm} = \sqrt{\frac{1\pm q}{2}} | \Psi^{\pm}\rangle \langle 10 |, & K_{4,\pm} = \sqrt{\frac{1\pm q}{2}} | \Phi^{\pm}\rangle \langle 11 |,
\end{eqnarray}
with the 8 Kraus operators satisfying $\sum_{i,\pm} K_{i,\pm}^{\dagger} K_{i,\pm} = \mathbb{I}$, where $\{ |\Phi^{+}\rangle ,|\Phi^{-}\rangle ,|\Psi^{+}\rangle ,|\Psi^{-}\rangle\}$ are the four pure Bell states defined in Eq.~(\ref{eq:BellStates}). The global rephasing map $\Lambda_{{\rm GR}_3}^{q}$, which is a physical CPTP map for all $q \in [-1,1]$, transforms any two-qubit state into a BD state lying on the phase flip freezing surface, i.e., $\Lambda_{{\rm GR}_3}^{q}(\rho) = \rho^q$, where $\rho^q$ is a BD state with characteristic triple given by $\{q,-q T_{33},T_{33}\}$, with $T_{33}=\mbox{Tr}\left[ (\sigma_{3} \otimes \sigma_{3})\rho \right]$. Specifically, the action of $\Lambda_{{\rm GR}_3}^{q}$ on a BD state $\rho$ which already belongs to the phase flip freezing surface is
\begin{equation}\label{eq:magicjohnson}
\Lambda_{{\rm GR}_3}^{q}(\{c_{1},-c_{1} c_{3},c_{3}\})=\{q,-q c_{3},c_{3}\}\,.
 \end{equation}
 This map is therefore able to restore the lost coherence for any (even completely dephased) BD state on the freezing surface, thus effectively reverting their decoherence process.

We  have then the inequality
\begin{eqnarray}\label{eq:inequalityduetorephasing1}
D(\{c_1,-c_1 c_3,c_3\},\{c_1,0,0\}) &=& D(\Lambda_{{\rm GR}_3}^{c_{1}}\{0,0,c_3\},\Lambda_{{\rm GR}_3}^{c_{1}}\{0,0,0\})  \nonumber \\
&\leq& D(\{0,0,c_3\},\{0,0,0\}),
\end{eqnarray}
where the first equality is due to the fact that
\begin{equation}
\{c_1,-c_1 c_3,c_3\} = \Lambda_{{\rm GR}_3}^{c_{1}} \{0,0,c_3\},\quad
\{c_1,0,0\} = \Lambda_{{\rm GR}_3}^{c_{1}}\{0,0,0\},
\end{equation}
and the final inequality is again due to contractivity of the distance $D$. By putting together the two opposite inequalities (\ref{eq:inequalityduetodephasing1}) and (\ref{eq:inequalityduetorephasing1}), we immediately get the invariance of  Eq.~(\ref{eq:magictranslationalinvariance1}) for any contractive distance.

To prove now the claim of Eq.~(\ref{eq:magictranslationalinvariance2}),  we introduce the unitary $U=\frac{1}{2}(\mathbb{I}+i\sigma_y)\otimes(\mathbb{I}+i\sigma_y)$ such that
\begin{equation}
U\{c_1,-c_1c_3,c_3\}U^\dagger=\{c_3,-c_1c_3,c_1\}, \quad
U\{0,0,c_3\}U^\dagger=\{c_3,0,0\}.
\end{equation}
By exploiting the invariance under unitaries of any contractive distance, and the just proven invariance expressed by Eq.~(\ref{eq:magictranslationalinvariance1}), we finally have
\begin{eqnarray}
& &D(\{c_1,-c_1c_3,c_3\},\{0,0,c_3\}) \nonumber \\
&=& D(U\{c_1,-c_1c_3,c_3\}U^\dagger,U\{0,0,c_3\}U^\dagger) \nonumber \\
&=& D(\{c_3,-c_1c_3,c_1\},\{c_3,0,0\}) \nonumber \\
&=& D(\{0,0,c_1\},\{0,0,0\}) \nonumber \\
&=& D(U\{0,0,c_1\}U^\dagger,U\{0,0,0\}U^\dagger) \nonumber \\
&=&D(\{c_1,0,0\},\{0,0,0\}),
\end{eqnarray}
that establishes Eq.~(\ref{eq:magictranslationalinvariance2}), thus concluding the proof of the Theorem.
\begin{flushright}
$\blacksquare$
\end{flushright}


We now prove that (D.i), (D.ii), (D.iii) $\Rightarrow$ (F.i), that is, one of the closest classical states to a BD state of the form
$\{c_1,-c_1c_3,c_3\}$, with $|c_1|>|c_3|$, is always the BD classical state of the form $\{c_1,0,0\}$ for any bona fide distance as defined above. We divide this result in two main steps, represented by Theorem \ref{th:nearestCCstatetoaBDstate} and Theorem \ref{th:theexactformoftheclosestCCstatetorho}. To bridge between the two Theorems we need  four Lemmas which are formulated and proved in the {\bf Methods}.

Let us begin by the following powerful result, which applies to all two-qubit BD states.
\begin{Theorem}\label{th:nearestCCstatetoaBDstate}
According to any contractive, transposition invariant, and convex distance, one of the closest classical states $\chi_\rho$ to a BD state $\rho$ is always a BD classical
state,
\begin{equation}\label{eq:theclosestBDclassicalstatetoaBDstate}
\chi_\rho = \frac{1}{4}\left(\mathbb{I}^A\otimes\mathbb{I}^B + s
\sigma_k^A\otimes\sigma_k^B \right)\,,
\end{equation}
for some index $k\in\{1,2,3\}$ and some coefficient $s\in[-1,1]$.
\end{Theorem}

\noindent \textit{Proof}. For an arbitrary two-qubit state $\varsigma$, described in the Bloch representation
\begin{equation}
\varsigma = \frac{1}{4}\left( \mathbb{I}^A\otimes\mathbb{I}^B + \sum_{i} x_{i} \sigma_{i}^{A} \otimes \mathbb{I}^{B} + \sum_{i} y_{i} \mathbb{I}^{A} \otimes \sigma_{i}^{B} + \\ \sum_{ij} T_{ij} \sigma_i^A\otimes\sigma_j^B \right),
\end{equation}
by the triple $\{\vec{x},\vec{y},\textbf{T}\}$, there exists another two-qubit state $\varsigma_{-}$ with the associated triple $\{-\vec{x},-\vec{y},\textbf{T}\}$ \cite{Aaronson2013a,Aaronson2013,Bromley2014}. The specific triple $\{\vec{x},\vec{y},\textbf{T}\}$ for a two-qubit BD state as defined in Eq.~(\ref{BDstate}) is $\vec{x}=\vec{y}=\vec{0}$ and $T_{ij}=\delta_{ij}c_i$; notice hence  that $\rho_- = \rho$ for BD states.

Given any BD state $\rho$, and any general two-qubit state $\varsigma$, it holds that
\begin{equation}\label{eq:danti}
D(\rho,\varsigma_{-})=D(O\rho O^{-1},O\varsigma O^{-1})=D(\rho,\varsigma),
\end{equation}
where we have introduced an antiunitary operator $O$ which maps $\varsigma$ to $\varsigma_-$. Explicitly, it acts in the computational basis as $O=(\sigma_y\otimes\sigma_y) C $, where $C$ denotes complex conjugation in the standard basis, which amounts to transposition for quantum states. Eq.~(\ref{eq:danti}) thus follows by exploiting the invariance of $D$ under antiunitary operations which is provided by properties (D.i) and (D.ii).

By further using the convexity of the distance $D$ in the second argument,
\begin{equation}
D(\rho,q \tau+(1-q)\varsigma) \leq q D(\rho,\tau) +(1-q) D (\rho,\varsigma),
\end{equation}
which is automatically implied by its joint convexity in property (D.iii), one can also show that for any BD state $\rho$ it holds that
\begin{equation}\label{eq:bensinequality}
D(\rho,\varsigma_0)\leq D(\rho,\varsigma),
\end{equation}
where $\varsigma_0$ is described in the Bloch representation by the triple $\{\vec{0},\vec{0},\textbf{T}\}$.
Explicitly,
\begin{equation}
D(\rho,\varsigma_0)=D\left(\rho,\frac{1}{2}\varsigma + \frac{1}{2}\varsigma_{-}\right)\leq \frac{1}{2}D(\rho,\varsigma)+\frac{1}{2}D(\rho,\varsigma_{-})= D(\rho,\varsigma)
\end{equation}
where we have used $\varsigma_0=\frac{1}{2}\varsigma + \frac{1}{2}\varsigma_{-}$ in the first equality, convexity of $D$ in the inequality and $D(\rho,\varsigma_-)= D(\rho,\varsigma)$ in the final equality.

We will now consider the distance from $\rho$ to the (larger) set of two-qubit CQ states, and show that its minimum can be attained by a fully classical (CC) BD state, hence proving the main result of the theorem. Recall that any CQ two-qubit state is of the form
\begin{equation}
\chi = p |\psi_1\rangle\langle\psi_1|^A\otimes\rho_1^B + (1-p)|\psi_2\rangle\langle\psi_2|^A\otimes\rho_2^B,
\end{equation}
where $p\in[0,1]$, $\{|\psi_1\rangle^A, |\psi_2\rangle^A\}$ is an orthonormal basis for qubit $A$ and $\rho_{1}^{B}$ and $\rho_{2}^{B}$ are arbitrary states of qubit $B$. Such a CQ state will have the associated triple $\{(2p-1)\vec{e},\vec{s}_+,\vec{e}\ \vec{s}^T_{-}\}$, where
\begin{equation}
e_i=\langle\psi_1|\sigma_i |\psi_1\rangle,\
s_{\pm,i}=\mbox{Tr}\left\lbrace\left[  p\rho_1 \pm (1-p)\rho_2\right] \sigma_i \right\rbrace ,
\end{equation}
with $\sigma_i$ being the Pauli matrices.
For any state in this form, a second state
\begin{equation}
\chi_0 = p' |\psi'_1\rangle\langle\psi'_1|^A\otimes{\rho'_1}^B + (1-p')|\psi'_2\rangle\langle\psi^{'}_2|^A\otimes{\rho'_2}^B,
\end{equation}
can be derived using the identities
\begin{eqnarray}
p'&=&\frac{1}{2}, \ |\psi'_1\rangle= |\psi_1\rangle,\  |\psi'_2\rangle = |\psi_2\rangle, \nonumber \\
{\rho'_1}^B&=&\frac{1}{2}\left[\mathbb{I}^B + p\tau_1^B - (1-p)\tau_2^B \right], \nonumber\\
{\rho'_2}^B&=&\frac{1}{2}\left[\mathbb{I}^B - p\tau_1^B + (1-p)\tau_2^B \right],
\end{eqnarray}
where $\tau_1$ and $\tau_2$ are the traceless part of $\rho_1$ and $\rho_2$. This state is manifestly CQ and it can be easily verified that it will have the associated triple $\{\vec{0},\vec{0},\textbf{T}\}$ with $\textbf{T}=\vec{e}\ \vec{s}^T_{-}$. From the inequality (\ref{eq:bensinequality}), we have in particular that $D(\rho,\chi_0)\leq D(\rho,\chi)$ for any BD state $\rho$ and any CQ state $\chi$,  so that in order to minimise $D(\rho,\chi)$ it suffices to restrict ourselves to CQ states with associated triple $\{\vec{0},\vec{0},\vec{e}\ \vec{s}^T_{-}\}$.

Temporarily, we relax the restriction that $\vec{e}$ is of unit length and consider the distance from the (even larger) set of states for which $||\vec{e}||\leq 1$. This is a convex set and so, due to the convexity of the distance, any local minimum will be a global one. We can now use a trick analogous to the one used for Eq.~(\ref{eq:danti}), this time between $\chi$ with $\vec{e}=(e_1,e_2,e_3)$ and $\vec{s}_{-}=(s_1,s_2,0)$ and $\chi'$ with $\vec{e}'=(e_1,e_2,-e_3)$ and $\vec{s}_{-}'=\vec{s}_{-}$. We then see that
\begin{equation}
D(\rho,\chi)=D(U_z\rho U_z^{\dagger},U_z\chi U_z^{\dagger})=D(\rho,\chi'),
\end{equation}
where we have introduced the unitary operator $U_z=\sigma_z\otimes\sigma_z$ such that $\rho=U_z\rho U_z^{\dagger}$ for any BD state $\rho$, $\chi'=U_z\chi U_z^{\dagger}$, and we have exploited (D.i). A similar result holds when considering distances from states with either $s_1=0$ or $s_2=0$, by using, respectively $U_x=\sigma_x\otimes\sigma_x$ and $U_y=\sigma_y\otimes\sigma_y$. Also, a similar result holds by switching the vectors we consider, for any $e_i=0$. From these observations we have that, if $s_i=0$ for some index $i$, then the minimum distance is attained for  $e_i=0$,  and viceversa.

We can then restrict our attention to states with $e_i=e\delta_{ik}$ and $s_{i}=s\delta_{ik}$, where the index $k$ sets the  nonzero vector element. From the previous results, we notice in fact that minimisation only needs to be performed over $e_k$ and $s_k$  as the distance can only decrease under any variation in any other single element. Furthermore, $e_k$ and $s_k$ appear only as a product $e_k s_k$ in the density matrix, never on their own. This means that minimising over both is equivalent to setting $e_k=1$ and minimising only over $s_k$, thus allowing us to reimpose the restriction that $||\vec{e}||=1$, thus coming back to analyse the distance from $\rho$ to CQ states. The remaining states over which the minimisation in the single parameter $s$ needs to be performed amount exactly to the set of BD classical states (aligned on the axes in Fig.~\ref{fig:thephaseflipfreezingsurface}), hence finding the minimum among these will return the global minimum for the distance $D$ from an arbitrary BD state $\rho$ to the set of two-qubit classical states, proving the claim.
\begin{flushright}
$\blacksquare$
\end{flushright}

What is left at this point is to perform the final minimisation, namely to determine exactly the values of the index $k$ and of the coefficient $s$ that entirely specify the closest classical state $\chi_\rho$ of Eq.~(\ref{eq:theclosestBDclassicalstatetoaBDstate}) as a function of the coefficients $\{c_1,c_2,c_3\}$ defining any given BD state $\rho$. When we restrict ourselves to BD states $\rho$ belonging to the phase flip freezing surface of Eq.~(\ref{eq:constraintdefiningthefreezingsurface}), the solution is provided by Theorem \ref{th:theexactformoftheclosestCCstatetorho}, which makes use of the auxiliary results proven in the {\bf Methods}.



\begin{Theorem}\label{th:theexactformoftheclosestCCstatetorho}
According to any convex and contractive distance, one of the closest classical states
$\chi_\rho$ to a BD state $\rho$ of the form $\{c_1,-c_1 c_3,c_3\}$ is
\begin{enumerate}
\item when $|c_1|\geq|c_3|$, the BD classical state $\{c_1,0,0\}$, i.e. the one with $k=1$ and $s=c_1$ in Eq.~(\ref{eq:theclosestBDclassicalstatetoaBDstate});
\item when $|c_3|\geq|c_1|$, the BD classical state
$\{0,0,c_3\}$, i.e. the one with $k=3$ and $s=c_3$ in Eq.~(\ref{eq:theclosestBDclassicalstatetoaBDstate}).
\end{enumerate}
\end{Theorem}

\noindent \textit{Proof}. According to Theorem \ref{th:nearestCCstatetoaBDstate}, one of the closest classical states to any BD state is a classical BD state. According to Lemmas \ref{lem:nearestclassicalBDstateinsidethexandzaxis} and \ref{lem:wecandiscardtheyaxis}, the closest classical BD state $\chi_\rho$ to a BD state $\rho$ of the form $\{c_1,-c_1 c_3,c_3\}$ is either $\{c_1,0,0\}$ or $\{0,0,c_3\}$. Finally, according to Lemmas  \ref{lem:equidistancefromtheoriginonthexandzaxis} and \ref{lem:contractionofthedistanceunderpuredephasingalongthexandzaxis}, if $|c_1|\geq|c_3|$ then one of the closest classical BD states to $\rho$ is $\{c_1,0,0\}$, whereas if $|c_3|\geq|c_1|$ then one of the closest classical BD states to $\rho$ is $\{0,0,c_3\}$. \begin{flushright}
$\blacksquare$
\end{flushright}

From a physical perspective, the most relevant implication of Theorem \ref{th:anycontractivedistanceismagictranslationalinvariant} and Theorem \ref{th:theexactformoftheclosestCCstatetorho} is that the freezing phenomenon, as described earlier adopting the guiding example of the Bures distance, occurs in fact for any bona fide distance-based measure of quantum correlations whose underlying distance is invariant under transposition, convex and contractive.

Namely, the condition $t<t^*$ is equivalent to $|c_1(0)e^{-2\gamma t}|>|c_3(0)|$, so that from Theorem \ref{th:theexactformoftheclosestCCstatetorho} we have that one of the closest classical states to the evolved BD state $\{c_1(0)e^{-2\gamma t},-c_1(0)c_3(0)e^{-2\gamma t},c_3(0)\}$ is $\{c_1(0)e^{-2\gamma t},0,0\}$ for any $t<t^*$. Therefore, referring to the definition of distance-based quantifier of discord-type quantum correlations given in Eq.~(\ref{eq:geometricquantumcorrelations}), one has
\begin{equation}
Q_D(\rho(t))=D(\{c_1(0)e^{-2\gamma t},-c_1(0)c_3(0)e^{-2\gamma t},c_3(0)\},\{c_1(0)e^{-2\gamma t},0,0\}),\ \ \forall t<t^*,\end{equation}
and according to Theorem \ref{th:anycontractivedistanceismagictranslationalinvariant} this is constant and equal to $D(\{0,0,c_3(0)\},\{0,0,0\})$ for any $t<t^*$.

On the other hand, the condition $t>t^*$ is equivalent to $|c_3(0)|>|c_1(0)e^{-2\gamma t}|$, so that from Theorem \ref{th:theexactformoftheclosestCCstatetorho} we have that one of the closest classical states to the evolved BD state $\{c_1(0)e^{-2\gamma t},-c_1(0)c_3(0)e^{-2\gamma t},c_3(0)\}$ is the BD state $\{0,0,c_3(0)\}$ for any $t>t^*$. Therefore, in this case the geometric quantifier of discord-type quantum correlations is
\begin{equation}
Q_D(\rho(t))=D(\{c_1(0)e^{-2\gamma t},-c_1(0)c_3(0)e^{-2\gamma t},c_3(0)\},\{0,0,c_3(0)\}),\ \ \forall t>t^*,\end{equation}
that, due to the contractivity of $D$, has to be monotonically nonincreasing for any $t>t^*$, eventually decaying to zero.

\section*{\sf \bfseries DISCUSSION}
In this paper we have established from first principles the general character of an intriguing dynamical trait of quantum correlations other than entanglement, namely their {\it freezing} under given environmental and initial conditions. This phenomenon manifests for the class of Bell-diagonal states of two qubits, which often constitute the simplest yet highly relevant class of states for which one is able to analytically calculate measures of quantum, classical, and total correlations, for instance by distance-based (geometric) quantifiers. In particular, we have shown here that a specific class of Bell-diagonal states of two qubits, each undergoing local non-dissipative decoherence, manifests freezing of discord-type quantum correlations whenever the distance adopted to measure them is assumed to be invariant under transposition, dynamically contractive and convex. As these physical properties are instrumental to define valid distance-based measures of correlations, our result means that freezing of quantum correlations occurs independently of the adopted distance and is therefore universal within a bona fide geometric approach.

Frozen quantum correlations have been verified both theoretically \cite{maziero, Mazzola,Mazzola2011,Aaronson2013a,PaulaEPL,necsuff, HRI} and experimentally  \cite{Xu2010,Auccaise2011,Cornelio2012,xulofranco2013NatComms,Silva2013,PaulaPRL} by using specific measures of quantum correlations \cite{Modi2012,Aaronson2013a}, but until now it was an open problem whether all suitable discord-type quantifiers (including potentially new ones yet to be defined) would freeze in the same dynamical conditions. Our work rigorously contributes to the settling of this problem and provides elegant evidence strongly supporting the conclusion that freezing of quantum correlations is a natural physical phenomenon and not merely a mathematical accident. Notice that freezing in BD states, as described in this paper, has also been observed for some discord-type measures which do not manifestly enjoy a distance-based definition, such as the local quantum uncertainty \cite{lqu,Aaronson2013a} and the interferometric power \cite{interpower}. This leaves some room for further research aimed to prove the occurrence of freezing from only the basic properties (Q.i), (Q.ii), and (Q.iii) of quantum correlations, possibly without the need to invoke a geometric approach as considered in this work. Alternatively, our result might suggest that all measures of discord could possibly be recast into a geometric form via some bona fide distance, at least when restricted to BD states of two qubits (this is the case, for instance, for the conventional entropic measure of discord \cite{Ollivier2001}, which becomes equivalent to the relative entropy-based discord \cite{Mazzola} for BD states); this would also be an interesting direction to explore, in a more mathematical context of information geometry.

We further remark that, although we have explicitly considered Markovian evolutions in our analysis, the freezing of quantum correlations also occurs in the presence of non-Markovian channels which can be described by a master equation with a memory kernel, as in the case of pure dephasing or decoherence under classical random external fields \cite{Mazzola2011,Haikka2013,LoFranco2012,LoFrancoAndersson2012,LoFranco2013}. Indeed, in these cases the dynamics of BD states can be formally written as in Eq.~(\ref{evolvc}), but with $2 \gamma t$ replaced by a more general time-dependent rate $\Gamma(t)$. This can give rise to a dynamics with multiple intervals of constant discord \cite{Mazzola2011,LoFranco2012,LoFrancoAndersson2012}, or discord frozen forever \cite{Haikka2013} depending on the initial conditions. By our analysis, we conclude that those fascinating features, which might be observable e.g.~in the dynamics of impurity atoms in Bose-Einstein condensates \cite{Haikka2013,gabriele}, are universal too and manifest when probed by any bona fide geometric discord-type measure $Q_D$.

Within this paper we introduced an intriguing global rephasing channel, which is able to reverse the effects of decoherence for certain two-qubit BD states. This physical CPTP channel may be of interest for applications other than proving the universality of the freezing phenomenon, for example quantum error correction \cite{nielsenchuang}, where it is desirable to combat the effects of noise, typically manifesting via local bit flip, phase flip, or bit-phase flip channels. For suitable BD states, all these errors can be corrected by global maps such as the one in Eq.~(\ref{Eq:KrausRepresentation}).  We also note that the action of this channel resembles (but is different from) the physical situation of refocusing by dynamical decoupling control on qubits undergoing low-frequency pure dephasing \cite{lofrancoAnnals,lofrancomataloni}. The further characterisation and experimental implementation of our global rephasing map for quantum information processing calls for an independent analysis which is beyond the scope of this paper.

From a fundamental perspective it is important to understand the deeper physical origin of frozen quantum correlations. There are reasons to reckon that the phenomenon is related to the complementary freezing of {\it classical} correlations. Typically, as observed so far using specific quantifiers, given  particular dynamical and initial conditions as studied here, quantum correlations are initially frozen and classical correlations decay but, after a characteristic time $t^*$, classical correlations freeze and quantum correlations decay \cite{Mazzola, Aaronson2013, Bromley2014}. This has been linked to the finite-time emergence of the classical pointer basis within the fundamental theory of decoherence \cite{Zurek1993,Cornelio2012,PaulaEPL,PaulaPRL}. Nevertheless, classical correlations are still inconsistently defined in geometric approaches  \cite{Bromley2014,SarandyCC} and it remains unknown whether they exhibit freezing after $t^\ast$ for any bona fide distance. This is certainly an aspect deserving further investigation.

Very recently, some of us have shown that an even more fundamental property of quantum systems, namely {\em coherence} \cite{coherence} in a reference basis, can also remain frozen under local nondissipative decoherence channels for the same class of initial states as studied here \cite{frozen}. Such a result holds more generally for a class of $N$-qubit states with maximally mixed marginals (for any even $N$), which include and extend the two-qubit set discussed in this work. This suggests that multiqubit and multipartite quantum correlations can freeze as well under the same dynamical conditions \cite{HRI,Xu2013}, and the methods of this work can be readily employed to prove the universality of  freezing within the geometric approach, in such a more general instance as well.

Our result has also an impact from an applicative point of view.  The property of being unaffected by the noise for a given period of time makes quantum correlations other than entanglement important for emergent quantum technologies \cite{merali,golden}. Despite numerous basic experimental investigations, this resilience has yet to be properly exploited as a resource for quantum enhanced protocols e.g.~in communication, computation, sensing and metrology. The universality of the freezing phenomenon for geometric quantum correlations, in paradigmatic quantum states and dynamical evolutions as shown here, promises to motivate further research in this context.


\section*{\sf \bfseries METHODS}

Here we derive some technical results needed for the proof of Theorem \ref{th:theexactformoftheclosestCCstatetorho}.

\begin{lemma}\label{lem:nearestclassicalBDstateinsidethexandzaxis}
According to any contractive distance $D$, it holds that:
\begin{enumerate}
\item among the BD classical states belonging to the $c_{1}$-axis, the closest state $\chi_\rho^{(1)}$ to a BD state $\rho$ of the form $\{c_1,-c_1c_3,c_3\}$ is the orthogonal projection of $\rho$ onto the $c_{1}$-axis, i.e. $\chi_\rho^{(1)}=\{c_1,0,0\}$;\label{item:nearestclassicalBDstateinsidethexaxis}
\item among the BD classical states belonging to the $c_{3}$-axis, the closest state $\chi_\rho^{(3)}$ to a BD state $\rho$ of the form $\{c_1,-c_1c_3,c_3\}$ is the orthogonal projection of $\rho$ onto the $c_{3}$-axis, i.e. $\chi_\rho^{(3)}=\{0,0,c_3\}$;\label{item:nearestclassicalBDstateinsidethezaxis}
\end{enumerate}
\end{lemma}
\noindent \textit{Proof}. Regarding point (\ref{item:nearestclassicalBDstateinsidethexaxis}), we need to prove that for any $x$
\begin{equation}
D(\{c_1,-c_1 c_3,c_3\},\{c_1,0,0\}) \leq D(\{c_1,-c_1 c_3,c_3\},\{c_1+x,0,0\}).
\end{equation}
In fact
\begin{eqnarray} \nonumber
& &D(\{c_1,-c_1 c_3,c_3\},\{c_1,0,0\}) \\ \nonumber
&=& D(\{0,0,c_3\},\{0,0,0\})   \\ \nonumber
&=& D(\Lambda^{\infty}_{{\rm LD}_3}\{c_1,-c_1 c_3,c_3\},\Lambda^{\infty}_{{\rm LD}_3}\{c_1+x,0,0\}) \\ \nonumber
&\leq& D(\{c_1,-c_1 c_3,c_3\},\{c_1+x,0,0\}), \nonumber
\end{eqnarray}
where the first equality is due to Theorem \ref{th:anycontractivedistanceismagictranslationalinvariant}, which holds for any contractive distance, the second equality is due to the fact that
\begin{equation}
\{0,0,c_3\} = \Lambda^{\infty}_{{\rm LD}_3} \{c_1,-c_1 c_3,c_3\},\
\{0,0,0\} = \Lambda^{\infty}_{{\rm LD}_3}\{c_1+x,0,0\},
\end{equation}
with $\Lambda^{\infty}_{{\rm LD}_3}$ representing complete local pure dephasing towards the $c_3$-axis, and finally the inequality is due to contractivity of the distance $D$.

Regarding point (\ref{item:nearestclassicalBDstateinsidethezaxis}) we need to prove that for any $z$
\begin{equation}
D(\{c_1,-c_1 c_3,c_3\},\{0,0,c_3\}) \leq D(\{c_1,-c_1 c_3,c_3\},\{0,0,c_3+z\}).
\end{equation}
In fact
\begin{eqnarray} \nonumber
& &D(\{c_1,-c_1 c_3,c_3\},\{0,0,c_3\})  \\ \nonumber
&=& D(\{c_1,0,0\},\{0,0,0\})  \\ \nonumber
&=& D(\Lambda^{\infty}_{{\rm LD}_1}\{c_1,-c_1 c_3,c_3\},\Lambda^{\infty}_{{\rm LD}_1}\{0,0,c_3+z\})  \\ \nonumber
&\leq& D(\{c_1,-c_1 c_3,c_3\},\{0,0,c_3+z\}), \nonumber
\end{eqnarray}
where the first equality is due to Theorem \ref{th:anycontractivedistanceismagictranslationalinvariant}, which holds for any contractive distance, the second equality is due to the fact that
\begin{equation}
\{c_1,0,0\} = \Lambda^{\infty}_{{\rm LD}_1}\{c_1,-c_1 c_3,c_3\},\
\{0,0,0\} = \Lambda^{\infty}_{{\rm LD}_1}\{0,0,c_3+z\},
\end{equation}
with $\Lambda^{\infty}_{{\rm LD}_1}$ representing complete local pure dephasing towards the $c_1$-axis, and finally the inequality is due to contractivity of the distance $D$.\begin{flushright}
$\blacksquare$
\end{flushright}

An analogous result does not hold for the classical BD states lying on the $c_2$-axis. However, due to the following Lemma \ref{lem:wecandiscardtheyaxis}, we can discard the classical BD states on the $c_2$-axis in order to find out the closest classical BD state to a BD state lying on the phase flip freezing surface.

\begin{lemma}\label{lem:wecandiscardtheyaxis}
According to any contractive distance $D$, it holds that:
\begin{enumerate}
\item the BD classical state $\chi_\rho^{(1)} = \{c_1,0,0\}$ on the $c_1$-axis closest to the BD state $\rho = \{c_1,-c_1 c_3,c_3\}$ is closer to $\rho$ than any classical BD state belonging to the $c_2$-axis;\label{item:discardingyaxispart1}
\item  the BD classical state $\chi_\rho^{(3)} = \{0,0,c_3\}$ on the $c_3$-axis closest to the BD state $\rho = \{c_1,-c_1 c_3,c_3\}$ is closer to $\rho$ than any classical BD state belonging to the $c_2$-axis.\label{item:discardingyaxispart2}
\end{enumerate}
\end{lemma}
\noindent \textit{Proof}. Regarding point (\ref{item:discardingyaxispart1}), we need to prove that for any $y$
\begin{equation}
D(\{c_1,-c_1 c_3,c_3\},\{c_1,0,0\}) \leq D(\{c_1,-c_1 c_3,c_3\},\{0,y,0\}).
\end{equation}
In fact
\begin{eqnarray} \nonumber
& &D(\{c_1,-c_1 c_3,c_3\},\{c_1,0,0\}) \\ \nonumber
&=& D(\{0,0,c_3\},\{0,0,0\})  \\ \nonumber
&=& D(\Lambda^{\infty}_{{\rm LD}_3}\{c_1,-c_1 c_3,c_3\},\Lambda^{\infty}_{{\rm LD}_3}\{0,y,0\})\\ \nonumber
&\leq& D(\{c_1,-c_1 c_3,c_3\},\{0,y,0\}), \nonumber
\end{eqnarray}
where the first equality is due to Theorem \ref{th:anycontractivedistanceismagictranslationalinvariant}, which holds for any contractive distance, the second equality is due to the fact that
\begin{equation}
\{0,0,c_3\} = \Lambda^{\infty}_{{\rm LD}_3}\{c_1,-c_1 c_3,c_3\},\
\{0,0,0\} = \Lambda^{\infty}_{{\rm LD}_3}\{0,y,0\},
\end{equation}
with $\Lambda^{\infty}_{{\rm LD}_3}$ representing complete local pure dephasing towards the $c_3$-axis, and finally the inequality is due to contractivity of the distance $D$.

Regarding point (\ref{item:discardingyaxispart2}), we need to prove that
\begin{equation}
D(\{c_1,-c_1 c_3,c_3\},\{0,0,c_3\}) \leq D(\{c_1,-c_1 c_3,c_3\},\{0,y,0\}).
\end{equation}
In fact
\begin{eqnarray} \nonumber
& &D(\{c_1,-c_1 c_3,c_3\},\{0,0,c_3\}) \\ \nonumber
&=& D(\{c_1,0,0\},\{0,0,0\})  \\ \nonumber
&=& D(\Lambda^{\infty}_{{\rm LD}_1}\{c_1,-c_1 c_3,c_3\},\Lambda^{\infty}_{{\rm LD}_1}\{0,y,0\})  \\ \nonumber
&\leq& D(\{c_1,-c_1 c_3,c_3\},\{0,y,0\}), \nonumber
\end{eqnarray}
where the first equality is due to Theorem \ref{th:anycontractivedistanceismagictranslationalinvariant}, the second equality is due to the fact that
\begin{equation}
\{c_1,0,0\} = \Lambda^{\infty}_{{\rm LD}_1}\{c_1,-c_1 c_3,c_3\},\\
\{0,0,0\} = \Lambda^{\infty}_{{\rm LD}_1}\{0,y,0\},
\end{equation}
with $\Lambda^{\infty}_{{\rm LD}_1}$ representing complete local pure dephasing towards the $c_1$-axis, and finally the inequality is due to contractivity of the distance $D$.\begin{flushright}
$\blacksquare$
\end{flushright}

\begin{lemma}\label{lem:equidistancefromtheoriginonthexandzaxis}
According to any contractive distance $D$, if $|c_1|=|c_3|$ then
\begin{equation}
D(\{c_1,0,0\},\{0,0,0\}) = D(\{0,0,c_3\},\{0,0,0\}).
\end{equation}
\end{lemma}
\noindent \textit{Proof}. Let us suppose that $c_1=\pm c_3=h$, then we have
\begin{eqnarray} \nonumber
& &D(\{h,0,0\},\{0,0,0\})  \\ \nonumber
&=& D(U_{\pm}\{0,0,\pm h\}U_{\pm}^\dagger,U_{\pm} \{0,0,0\}U_{\pm}^\dagger)  \\ \nonumber
&=& D(\{0,0,\pm h\},\{0,0,0\}), \nonumber
\end{eqnarray}
where the first equality is due to the fact that
\begin{equation}
\{h,0,0\} = U_{\pm}\{0,0,\pm h\}U_{\pm}^\dagger,\
\{0,0,0\} = U_{\pm}\{0,0,0\}U_{\pm}^\dagger,
\end{equation}
with $U_{+}=\frac{1}{2}(\mathbb{I}+i\sigma_y)\otimes(\mathbb{I}+i\sigma_y)$ and $U_{-}=\frac{1}{2}(\sigma_y+i\mathbb{I})\otimes(\mathbb{I}+i\sigma_y)$ being unitaries, whereas the second equality is due to unitary invariance of any contractive distance $D$.\begin{flushright}
$\blacksquare$
\end{flushright}

\begin{lemma}\label{lem:contractionofthedistanceunderpuredephasingalongthexandzaxis}
According to any contractive distance $D$, for any $q\in[0,1]$ the following holds:
\begin{eqnarray}\label{eq:contractionofthedistanceunderpuredephasingalongx}
D(\{qc_1,0,0\},\{0,0,0\}) \leq D(\{c_1,0,0\},\{0,0,0\}), \\
D(\{0,0,qc_3\},\{0,0,0\}) \leq D(\{0,0,c_3\},\{0,0,0\}).\label{eq:contractionofthedistanceunderpuredephasingalongz}
\end{eqnarray}
\end{lemma}
\noindent \textit{Proof}. Regarding Eq.~(\ref{eq:contractionofthedistanceunderpuredephasingalongx}), we have
\begin{eqnarray} \nonumber
D(\{qc_1,0,0\},\{0,0,0\}) &=& D(\Lambda_{{\rm LD}_3}^{t_q}\{c_1,0,0\},\Lambda_{{\rm LD}_3}^{t_q}\{0,0,0\}) \nonumber \\
&\leq& D(\{c_1,0,0\}),\{0,0,0\}), \nonumber
\end{eqnarray}
where the equality is due to the fact that
\begin{equation}
\{qc_1,0,0\} = \Lambda_{{\rm LD}_3}^{t_q}\{c_1,0,0\} , \,\,\,\,\,\,\,
\{0,0,0\} = \Lambda_{{\rm LD}_3}^{t_q}\{0,0,0\},
\end{equation}
with $\Lambda_{{\rm LD}_3}^{t_q}$ representing local pure dephasing towards the $c_3$-axis until the time $t_q$ such that $e^{-2\gamma t_q}=q$, whereas the inequality is due to contractivity.

Regarding Eq.~(\ref{eq:contractionofthedistanceunderpuredephasingalongz}) we have
\begin{eqnarray} \nonumber
D(\{0,0,qc_3\},\{0,0,0\}) &=& D(\Lambda_{{\rm LD}_1}^{t_q}\{0,0,c_3\},\Lambda_{{\rm LD}_1}^{t_q}\{0,0,0\})  \nonumber \\
&\leq&  D(\{0,0,c_3\}),\{0,0,0\}) \nonumber
\end{eqnarray}
where the equality is due to the fact that
\begin{equation}
\{0,0,qc_3\} = \Lambda_{{\rm LD}_1}^{t_q}\{0,0,c_3\},\
\{0,0,0\} = \Lambda_{{\rm LD}_1}^{t_q}\{0,0,0\},
\end{equation}
with $\Lambda_{{\rm LD}_1}^{t_q}$ representing local pure dephasing towards the $c_1$-axis until the time $t_q$ such that $e^{-2\gamma t_q}=q$, whereas the inequality is due to contractivity.\begin{flushright}
$\blacksquare$
\end{flushright}

\section*{\sf \bfseries ACKNOWLEDGMENTS}
We acknowledge discussions with I. Bengtsson, L. A. Correa, S. Di Martino, J. Filgueiras, F. Illuminati, G. Marmo, M. Piani, F. Plastina, D. \v{S}afr\'{a}nek, R. Serra, J. Settino, I. A. Silva, D. O. Soares-Pinto and K. \.Zyczkowski. GA acknowledges the kind hospitality and fruitful interaction with members of the QuIC group at the Harish-Chandra Research Institute (India), where this paper was finalised. This work was supported by the Brazilian funding agency CAPES (Pesquisador Visitante Especial Grant No.~108/2012), the Foundational Questions Institute (Physics of Information Grant No. FQXi-RFP3-1317), the ERC StG GQCOP (Grant No.~637352), and the FP7 European Strep Projects iQIT (Grant Agreement No.~270843) and EQuaM (Grant Agreement No.~323714).



\end{document}